# Towards slime mould chemical sensor: Mapping chemical inputs onto electrical potential dynamics of *Physarum Polycephalum*

James G.H. Whiting[1], Ben P.J. de Lacy Costello [1,2] Andrew Adamatzky[1]

[1] Unconventional Computing Centre; University of the West of England; Bristol, UK,
[2] Institute of Biosensing Technology; University of the West of England; Bristol, UK

**ABSTRACT**

Plasmodium of slime mould *Physarum polycephalum* is a large single celled organism visible unaided by the eye. This slime mould is capable of optimising the shape of its protoplasmic networks in spatial configurations of attractants and repellents. Such adaptive behaviour can be interpreted as computation. When exposed to attractants and repellents, Physarum changes patterns of its electrical activity. We experimentally derived a unique one-to-one mapping between a range of selected bioactive chemicals and patterns of oscillations of the slime mould's extracellular electrical potential. This direct and rapid change demonstrates detection of these chemicals in a similar manner to a biological contactless chemical sensor. We believe results could be used in future designs of slime mould based chemical sensors and computers.

**1. Introduction**

Cell based biosensors have been developed for several decades, they differ from traditional sensors as they use a cell or cell constituent as the sensing elements or transducers [24], with a range of applications from toxicity studies to environmental chemical sensing, a large majority of the cells used in this application are bacterial, due to the ease of genetic manipulation and the range of substrates they can detect; Other cells for biosensors are yeast or fungi based, which offer distinct advantages over bacterial based sensors [25]. Conditioning or genetic modification of cells has been demonstrated by several groups where specific genes and responses may be invoked. Despite the large volume of research into bacterial bio-sensors, very few have been developed commercially, mainly due to the fragility and short life of said sensors; bacteria also have limited temperature, pH conditions in which they will survive and will often not grow on specific substrates which would be ideal for cell-transducer interface. Yeast and wild fungi sensors are a lot more robust and grow in a larger variety of conditions while also offering advantages such as high growth rate and the ability to grow on a large range of surface substrates [26]. A significant benefit of using yeast is the very long shelf life of yeast cells, which can survive for over a year after dehydration and could be rehydrated when sensor use is required.

Another eukaryote, *Physarum polycephalum* has been shown to proliferate on a large range of surfaces such as plastics, agar, metals and glass; any number of which could form part of a transducer-cell interface. Drying out *Physarum polycephalum* will also produce a long lasting sclerotium which can be revived with moisture into a healthy plasmodium, in a similar manner to drying and rehydrating yeasts, which again is advantageous over bacterial cells for longevity and shelf life.

Myxomycetes, commonly known as Slime Moulds, are unicellular organisms belonging to the *Amoebozoa* kingdom; one such slime mould is *Physarum polycephalum*, which, in the plasmodial phase of its life cycle consists of a large single celled mass of yellow plasmodium. The organism extends protoplasmic tubes which grow towards sources of food; flowing through these tubes is a cytoplasm, which oscillates back and forth by a process of protoplasmic streaming which forces the cytoplasm in the direction in which the organism is growing [1]. The movement and growth of *P. polycephalum* is predominantly controlled by favourable conditions, such as an abundance of food, warm temperature, darkness and moisture. It has been well documented that various substances trigger a chemotactic response in *P. polycephalum*, various carbohydrates such as glucose and maltose initiate positive chemotaxis while sucrose shows marginal but consistent negative chemotaxis [2, 3, 4, 5].

Many other stimuli are known to provoke a response in *P. polycephalum*, for example, light exposure shows a phototactic response termed photoavoidance, with the organism moving away from sources of both white [6], blue light [7] and ultra-violet light [8]. In the 1999 paper, Nakagaki also showed that the frequency of oscillation could be phase shifted and frequency locked to rhythmic pulses of white light. Temperature is another stimulus, with literature suggesting Physarum prefers warm conditions, showing growth and migration from $18^oC$ towards $35^oC$ when a temperature gradient was applied to the supporting agar medium [4]. The most obvious and probably most well documented stimuli of Physarum is food sources with commercially available oat flakes being employed as a common source of nutrients when culturing Physarum. It is believed that carbohydrates provide a food source for the organism, with several sugars showing strong chemotactic attraction; other chemicals have been tested such as simple volatile organic chemicals (VOCs) [9] by way of binary choice experiments, with farnesene, b-myrcene, tridecane and other molecules producing chemo-attraction, while benzyl-Alcohol, geraniol and 2-phenylethanol among others produced chemo-repulsion. It is not thought that these VOCs are sources of food, however it was suggested that oxygen functionality and cAMP inhibition play a role in the chemicals' chemotactic outcome; as a result of this paper, it is evident that while Physarum is attracted to food sources, it is also attracted or repelled by chemicals which may not be food sources. It is possible that these chemicals either have a direct effect on non-specific membrane receptors or that there are until-now unknown behavioural stimuli whereby Physarum responds to chemicals which naturally occur as pheromones and secretions of organisms within the same ecological environment, adopting the protection from other lifeforms or possibly as a method of locating larger food sources rather than small detritus.

Various studies have shown Physarum's protoplasmic movement is based around an oscillatory system known as shuttle streaming with an inherent period of from 1 to 5 minutes; the exposure to both positive and negative chemotactic agents increase and decrease the frequency respectively. This concept has been observed when measuring the membrane under the microscope, visually quantifying the change in the protoplasmic tube with the associated oscillation, however measurements have been made on the electric potential of the protoplasmic tubes during oscillation [10], with Kashimoto [11] documenting an average surface potential of -83.5 mV, with a normal oscillation of approximately 5 mV with a period of 1.5 to 2 minutes, a frequency similar to that observed optically. It is understood that the oscillation varies between experimental set ups and life cycle state [12] with a general amplitude of between 5 to 10 mV and a period of 50 to 200 seconds, which has been proposed is the change in potential due to shuttle streaming movement. These experiments were largely performed with short protoplasmic tubes fixed against electrodes, and some papers suggest that although the frequency is similar, the electrical potential oscillation may be independent of shuttle streaming oscillation. This paper hopes to validate or reject a correlation between the shuttle streaming and electrical oscillations.

Computation using Physarum is an emerging field, in relatively early stages, with computation based around the observed laws governing the organism's natural instinct to hunt for food sources [13], with more complex implementations using a combination of long and short range, attractant and repellent chemicals. One existing limitation of Physarum computation is the time it takes for the plasmodium to extend protoplasmic tubes towards the food particles; other limitations exist which are more practical, such as keeping the organism alive and maintaining viable environmental conditions, as well as developing a reusable rather than single use computer [14].

All present computers take advantage of the spatial awareness of the organism, with the majority of Physarum computers solving geometric problems based on shortest-path approximation [15, 16, 17, 18, 19] or multi organism interaction [20], however some have been designed which allow the processing of other computation tasks such as simple logic gates. These computers require visual inspection of the organism either with the naked eye, or time lapse photography, however it would be ideal to be able to inspect the growth of the plasmodium with automated processing, outputting answers without human intervention, or even measuring the electrical potential of a multi-electrode array to calculate the growth and movement of Physarum to form a closed loop computer.

Following the plan of our EU funded project PhyChip [21] we aim to design and fabricate a distributed biomorphic computing device built and operated by slime mould *P. polycephalum*. A Physarum chip is a network of processing elements made of the slime mould's protoplasmic tubes coated with conductive substances; the network is populated by living slime mould. A living network of protoplasmic tubes acts as an active nonlinear transducer of information, while templates of tubes coated with conductor act as fast information channels. The proposed Physarum chip will require myriad of control and data inputs: optical, chemical, mechanical and electrical. In [22, 23] we experimentally implemented a tactile sensor with *P. polycephalum*. In the present paper we evaluate the feasibility of implementing a tactile sensor from the slime mould.

Preconditioning of cells in a biosensor can produce reliable and repeatable results for predetermined and specific chemicals or metal ions [27], as they have high bioabsorption for metal ions, however if a specific trait such as bioabsorbtion is not present, genetically modified fungi or bacteria can be introduced with desirable traits. While genetic modification (GM) can be easily performed in bacteria with the addition of specific plasmids, Autonomously-Replicating Sequences (ARS) containing shuttle vectors have very limited use with filamentous fungi making genetic modification more complex but still achievable. Currently no permanent genetic modification has been performed on Physarum polycephalum as only transient electroporated genetic information transfer has been managed thus far [28]. The growth conditions for Physarum are also very broad, and can grow in a variety of naturally changing and diverse conditions, leading themselves towards both bio-sensors and Biochemical Oxygen Demand (BOD) sensors. BOD sensors are used to evaluate the effectiveness of wastewater treatment plants and industrial factory water outlets, whose wastewater has the ability to inhibit or suppress microorganism growth and proliferation; bacterial and fungal bio-sensors using integrated cells to measure BOD have been developed [25], with reliable and repeatable current based output. Other yeast biosensors detect catabolic substrates which measure amounts of said substrates in solution, however these are subject to multiple substrate interference which can cause issues in such environments. *Hansenula anomala* immobilised on a pH sensor, allows for the detection of glucose and other simple saccharides whereby the change in pH of the organism with the presence of the carbohydrates is detected by the pH sensor [29], an application of which enables the detection of glucose in blood, however whether it would be sensitive enough to act as a diabetic monitor remains unclear; others have detected lactate in blood using a similar set up. Other fungi have been used for environmental and bioremediation purposes, with fungi detecting diesel oil in contaminated soil samples, facilitating the monitoring and evaluation of oil spill sites. Biosensors are also being developed for use to detect toxic agents chemical agents [30], employing intra-cell components such as enzymes or receptors as the biological sensing section. Other iterations of biosensors detect cancerous cell markers [31] and food spoilage [32, 33]

Yeasts and filamentous fungi have been genetically modified to express bioluminescence or fluorescence [25] for bioassays detecting hazardous or illegal substances; other genetically modified bacterial and fungal based sensors evaluate the toxicity and genotoxicity of antibacterial and antifungal substances, helping to evaluate sanitary agents in sterile environments. GM yeast is being developed into *in vitro* assessments for drug and chemical tests, which better mimic human cells than bacteria; Physarum is another eukaryotic organism, like fungi, so has the potential to be developed into a cell-based sensor for *in vitro* applications. It has also been reported that a biosensor with the same species of GM fungus has shown a more accurate parallel for human toxicity than the Salmonella based Ames test.

One important part of a cell based bio-sensor is the cell incorporation and signal detection, with cells suspended in solution, entrapped in porous membranes or immobilised on transducer surfaces, the biocompatibility between cell and sensor interface is key; another strength of Physarum is, as mentioned, its ability to grow on almost any surface. A common method of signal detection from bacteria or fungi is the Clark amperometric method which measures the current flow across a surface or solution between reference and recording electrodes, alternatively automated flow cells may be used, however with the bioluminescent cells, simple luminescence meters may be employed. Those cells which change pH or $CO_2$ conditions use dedicated transducers to indirectly measure the cellular activity.

Hulaniki et al. simply defined a chemical sensor as a device that transforms a chemical input into an analytically useful signal [34], it is believed that *Physarum polycephalum* fits this definition. The slime mould's inherent oscillation changes when exposed to a plurality of chemicals [2, 4, 5], which can be measured electronically; the measured change in frequency and amplitude equates to analytically useful signal. While most sensor technology currently being developed appears to be in the form of Microelectromechanical systems (MEMS) [36] or gas sensing nano structures [37], there is an emergent area of research which investigates biologically inspired or biologically integrated chemical sensors.

Biological sensors are often far more sensitive than manufactured chemical sensors, for example, a trained dog is still used ahead of chemical sensors to detect drugs or explosives in security situations; biologically inspired sensors are reported to be similar in sensitivity as these biological systems, [38]. Bacteria are also employed as chemical sensors [39], proving that while there is manufactured integration required for signal acquisition, the transducer portion of a chemical sensor may be organism based.

It is the aim of this research to develop an understanding of the electrical response to chemical stimuli with a view to providing such a closed loop electrical computer; a range of chemicals with known chemotactic responses are tested and, where possible, the results compared to those reported when optically measuring the change in frequency after exposure to chemicals. The advantage of logging the electrical potential of *Physarum polycephalum* is evident, providing continuous data without the need for complex visual processing, this paper aims to advance the field of Physarum computers by investigating and documenting for the first time, the electrical potential response to the addition of a variety of chemicals.

## 2. Materials and Methods

*2.1. Culturing Physarum polycephalum*

The plasmodium of *P. polycephalum* was grown using two methods; one technique was employed in order to maintain a pure culture of Physarum, using a plastic tub lined with damp kitchen paper, refreshed weekly, fed periodically with oat flakes. The other method was used to provide individual oat flakes inoculated with Physarum, which facilitated the transplantation of Physarum with a short term food source onto the experimental set up described below; this method used non-nutrient 2% agar in sterile 9 cm Petri-dishes and were fed daily with a small amount of oat flakes, to encourage growth. The latter method was repeated as required, when the Physarum had grown to the size of the Petri-dish and exhausted the supplied nutrient source. Both cultures of *P. polycephalum* were kept in a dark store at 20 $^{\circ}$C, removed into the light, only to extract the slime mould samples on oat flakes for experiments, upon completion of which they were returned to the dark store.

*2.2. Apparatus for measuring electrical potential of Physarum polycephalum*

To facilitate the automatic continuous recording of the voltage along a protoplasmic tube of *P. polycephalum*, both before and after the addition of chemical, a voltage data acquisition system was required. A laptop installed with Windows Vista was employed in collaboration with a PicoLog ADC-24 High resolution data logger, with 16 channels and 24 bit resolution. (Pico Technology, UK). The PicoLog was connected to the laptop using a USB connection, streaming converted digital data to the laptop using the associated *PicoLog Recorder* V5.22.8 software, the input channels were set to +/- 39mV ground referenced recording and a sampling frequency of 2 Hz; simultaneous channels were recorded as required.

The 9 cm Petri dishes (Fisher Scientific, UK) were customised to facilitate the recording of a single protoplasmic tube; two lengths of electrically conductive shielding tape (Advance Tape AT521, RS Components, UK) were placed at the centre of the dish on opposing sides, with approximately a 10 mm gap, and the length of the tape extended outside the Petri dish so crocodile clips could be attached. At the end of each section of each tape, in the centre of the dish, 1ml of 2% agar was placed; upon one a bare organic rolled oat flake was placed, and on the other blob, an organic oat flake which had been inoculated with *P. polycephalum* was placed, the idea being the plasmodium on the inoculated oat flake, after digesting it, would grow a tube towards the fresh oat flake, leaving a single protoplasmic tube between the two recording electrodes. The experimental set up is shown in Fig. 1 (a). The assembled Petri dishes were placed in a dark store, with their lids on, and the slime mould allowed to grow until a single tube had grown between the two oat flake topped agar blobs, with both agar blobs and oat flakes being colonised fully by the organism (Fig. 1 (b) and Fig. 1 (c)). Upon successful completion of growing, the original inoculated oat flake was connected to analogue ground on the PicoLog and the newly inoculated oat flake connected to an analogue recording channel. The method was first proposed and successfully tested in [21, 22, 40], measuring the surface potential difference of the Physarum protoplasmic tube which extends between the agar blobs; growing the blobs on agar causes a lower voltage than initially suggested [10, 11] due to the increased resistance of the agar, however the agar blobs maintain normal behaviour and extend its life beyond that grown purely on plastic or metal.

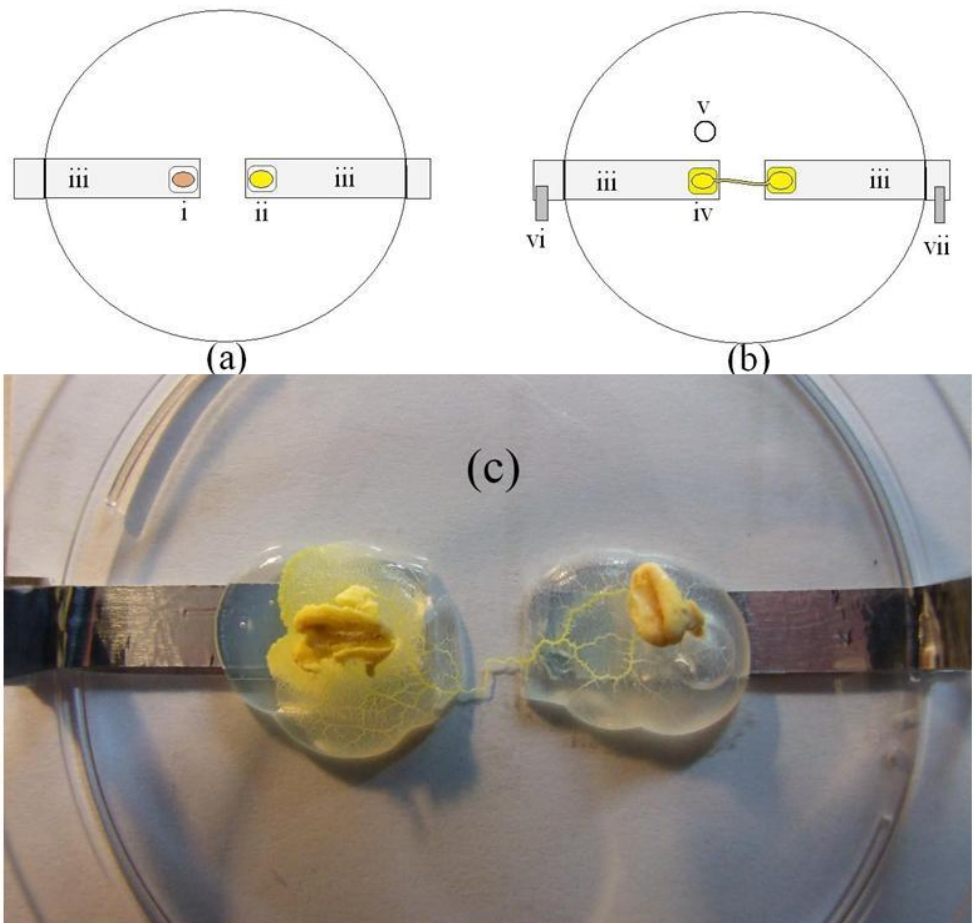

Figure 1. (a) Petri dish set up for chemotactic assessment before growth. (b) Petri dish with correct growth of Physarum polycephalum, connected for electrical potential recording. (c) example of protoplasmic tube growth and 2% Agar blob. i Bare oat flake. ii Physarum inoculated oat flake. iii conductive tape. iv single protoplasmic tube between electrodes. v site of chemical addition. vi positive recording terminal. vii ground reference terminal.

*2.3. Chemicals for chemotactic testing*

The choice of chemicals was derived from previous literature, detailing a hierarchy of attractants and repellents for *P. polycephalum* [9], with a variety of types of chemicals chosen based on observations made by Costello et al. From this group of chemicals a sub-group were chosen to represent a spectrum of attractant and repellent power; in addition to these chemicals, varying concentrations of agar gel were tested as it was noticed during culturing that Agar grew more quickly towards blobs of 6% agar; concentrations of agar gel have been shown in previous literature to have similar *Physarum* supporting qualities as weaker solutions of glucose, so varying strengths of agar were tested for attractant power. A sample of the VOCs were chosen, as tested by de Lacy Costello and Adamatzky [9] where authors had detailed their attractant power by use of binary choice between two chemicals. The frequency changing power of simple carbohydrates such as glucose and sucrose is well documented using the microscope recording method [4]; these carbohydrates act as food sources for Physarum so it was decided to test the frequency changing power of non-food source chemicals.

*2.4. Experiment and analysis method*

When the experiment is set up as shown in Fig. 1 (b), the recording can commence; recording is made and visualised simultaneously on screen in order to observe the oscillation of the protoplasmic tube. Depending on the state of the migration or growth, various waveforms can be seen, however the natural oscillation associated with shuttle streaming is typically observed within half an hour of recording; after 5 periods of stable oscillation are recorded, the time is noted and chemical is added. The chemicals, with the exception of the varying concentrations of agar gel, are added by dipping a 5 millimetre circle of filter paper into the chemical until saturation, then placing this 10 millimetres from the recording electrode agar blob, in parallel with the tip, in order to provide enough attraction so that Physarum may grow from the ground electrode toward the recording electrode. The lid was in place throughout the experiment, only being lifted to place the chemical into the dish; the experiment was performed in dark conditions using the minimum amount of light when placing chemical. The recording was continued for a period of approximately 30 minutes after exposure to the chemical; a successful recording was obtained when at least 5 oscillations were seen shortly after the chemical, the recordings with immeasurable oscillations after chemical addition were discarded. This procedure was repeated until at 12 successful recordings were acquired for each chemical. The method for administering the agar gel was similar, however small 5 millimetre discs of gel approximately 2 millimetres high were used in place of the filter paper. Each data file was exported to Matlab 7.0.1 and Microsoft office Excel 2003 for data processing. The frequency of oscillations was measured peak to peak between periods; amplitude was measured from peak to trough, an example of this technique is shown in figure 2. Mean frequency before and after was calculated for each Petri-dish, with the relative change being calculated as the frequency after chemical divided by the frequency before chemical; with an answer greater than 1 indicating a decrease in frequency or an increase in period. Amplitude was processed in the same manner, with an outcome greater than 1 indicated a decrease in amplitude after chemical.

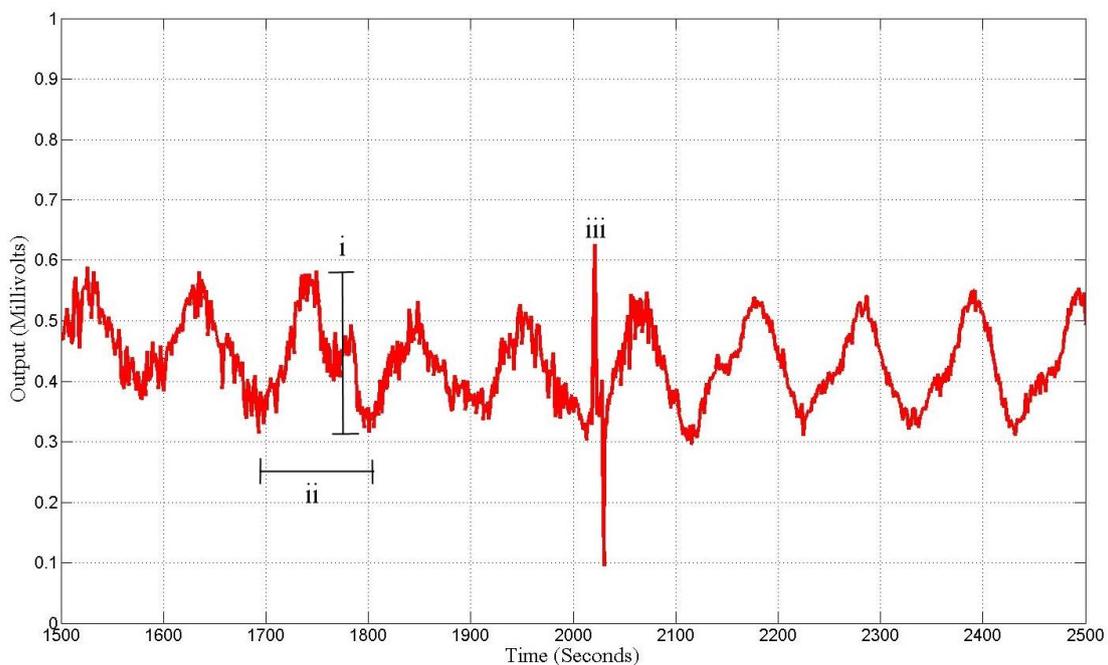

Figure 2. Example of recording before and after chemical, demonstrating amplitude (i) and frequency (ii) measurement, with mechanical stimulation spike (iii).

*2.5. Agar as an attractant*

A Physarum inoculated oat flake was placed on a circle of 20mm diameter filter paper at the centre of a Petri dish, onto which, tap water had been dropped until obvious saturation occurred. A blob of non-nutrient agar had been placed approximately 15mm away from the Physarum in the centre. 9 plates of 2% and 9 plates of 6% were produced for repeatability and statistical testing (18 in total). The Physarum was left for 2-3 days to allow for migration or growth towards the agar if it was to occur; the filter paper was topped up with water if needed, to avoid evaporation). In addition to this test, small 10 millimetre wide 2 millimetre thick discs of agar were produced, and were tested in a similar manner to a chemical, recording the electrical response of 2%, 4%, 6% and 8% non-nutrient agar gel.

**3. Results**

*3.1. Chemicals for chemotactic testing*

The efficiency of growth of Physarum into a suitable tube for the experimental set up (fig 1.b) was approximately half; where if 20 dishes were prepared, about 10 grew towards the bare oat flake with a single tube. Unsatisfactory growth was the result of multiple tube formation between the blobs, sclerotia formation, growth away from the oat flake which was often along the edge of the conductive tape suggesting that the organism was attracted to the glue underneath, or any other condition which did not enable recording. Even when growth was satisfactory, 5 suitable simultaneous oscillations did not always occur; 42 plates from total 240 single tube Petri dishes showed this. After chemical addition, a small number of Petri dishes did not provide measurable oscillations most often when large sporadic spikes appeared.; strong repellents such as Linalool and Nonanal evoked a less (than other chemicals) reliable response, as their addition was often swiftly followed by permanent oscillatory cessation, suggesting death of the organism. The addition of these two chemicals at further distances produced less of these terminal recordings, so for Linalool and Nonanal the distance from the recording agar blob was approximately 30 millimetres.

Table 1. Summary of mean frequency and amplitude changes for each chemical

| Chemical | Mean Frequency Change (Standard Deviation) | Mean Amplitude Change (Standard Deviation) |
|---|---|---|
| Farnesene | 1.255  (0.249) | 0.646  (0.167) |
| Tridecane | 1.170  (0.367) | 1.104  (0.457) |
| S(-)Limonene | 1.013  (0.064) | 0.970  (0.328) |
| Cis-3-Hexenylacetate | 0.987  (0.210) | 0.941  (0.462) |
| Geraniol | 0.999  (0.105) | 0.740  (0.220) |
| Benzyl Alcohol | 0.893  (0.114) | 1.259  (0.254) |
| Linalool | 0.676  (0.213) | 1.414  (0.314) |
| Nonanal | 0.731  (0.164) | 0.718  (0.186) |

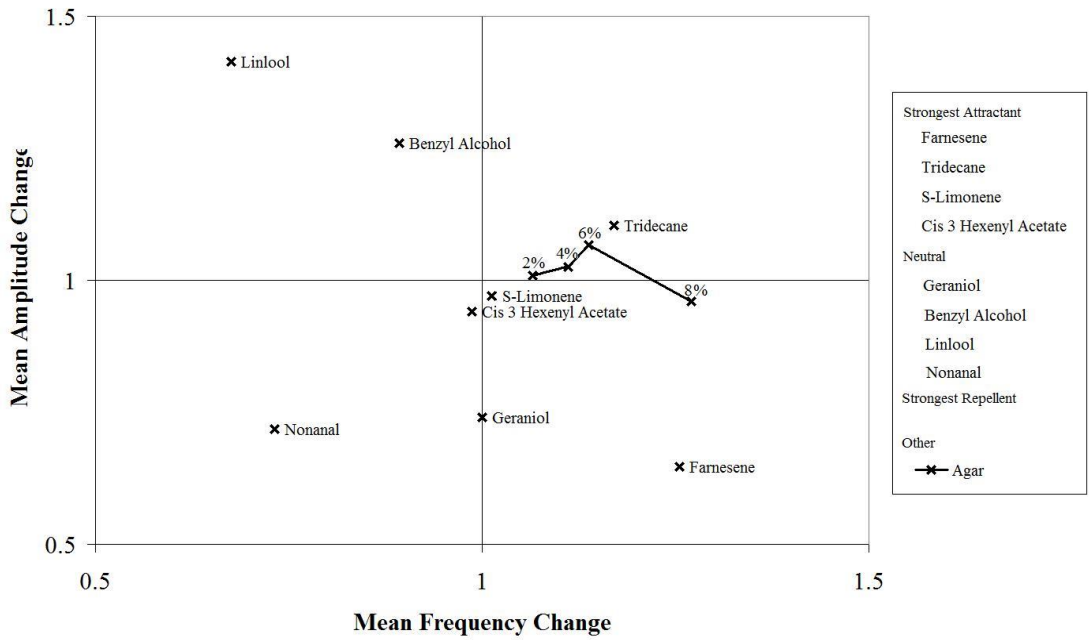

Figure 3. Summary of changes to frequency and amplitude after chemical exposure.

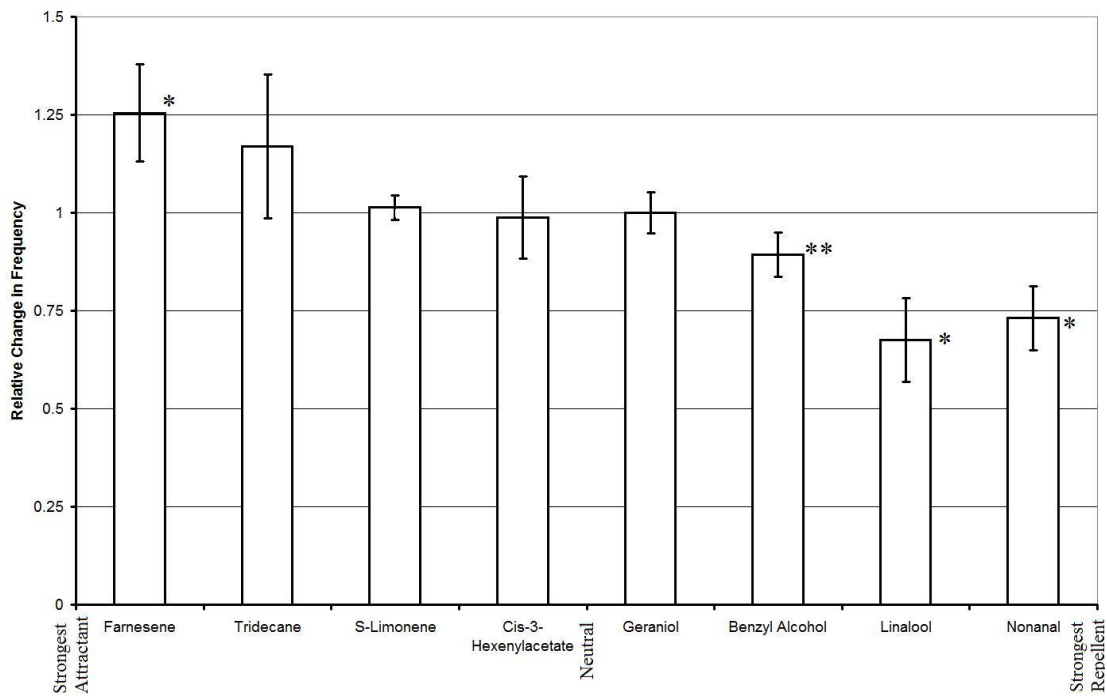

Figure 4. Relative frequency change ranked from strongest attractant to strongest repellent. * denotes statistical significance of $P < 0.05$, ** denotes statistical significance of $P < 0.1$.

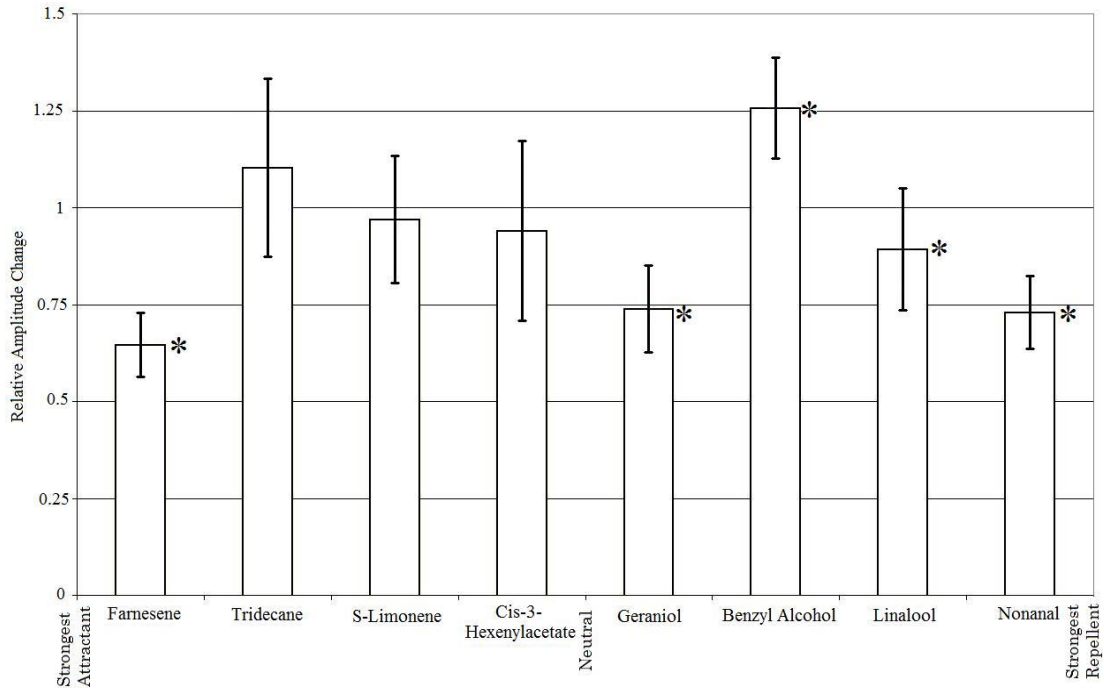

Figure 5. Relative amplitude change ranked from strongest attractant to strongest repellent.
* denotes statistical significance of P < 0.05.

*3.2. Agar as an attractant*

Following the author's observations that agar blobs of 6% can act as an attractant with no other nutrient content or oat flakes and that Physarum appears to grow more quickly across 6% agar then 2%, the attractant power of agar was investigated; a paper was found detailing the growth and migration of Physarum to carbohydrates including agar [2] which concluded that agar was acting as a carbohydrate source, with 2% non nutrient agar gel providing similar growth and migration rates as 0.05% glucose solution. This showed that agar itself could be a weak attractant, providing enough carbohydrate to support growth, even in the absence of other food sources. The 6% non-nutrient agar would be providing more nutrients and presumably be a stronger attractant than a weaker solution of agar. It had been suggested that the possible attractant component could be water content in the agar, so a varying concentrations of agar were tested using a custom set up described below, which would test this hypothesis; also varying levels of agar gel were included in the main study to test for their relative attractant power.

## 4. Discussion

The natural oscillating frequency of the *Physarum polycephalum* appears to match the frequency recorded optically [4, 6, 35, 41] and electronically [3], with a period ranging from 90 to 170 seconds before any chemical is added. This suggests that recording the electrical potential of a protoplasmic tube between two oat flakes on agar is a reliable method of measuring protoplasmic shuttle streaming.

The majority of chemicals tested follow the general rule that attractants increase frequency and repellents decrease frequency, although not every chemical follows this trend, shown in fig. 4; the variability between the same chemical shows that it is difficult to measure the frequency change in an individual experiment and that repeats are required to accurately assess the nature of the frequency change, this is highlighted as only the strong attractants and repellents show statistical significance. The overlap of chemicals appears in the neutral zone, with very weak attractants S(-)Limonene and Cis-3-Hexenyl Alcohol having almost no effect on either amplitude or frequency. Weak repellent Geraniol does not appear to affect frequency however has a notable reducing effect on amplitude suggesting the chemical may not induce growth but that the organism is still capable of detecting it, shown as a statistically significant reduction of the amplitude as seen in fig. 5. It is possible that the small mean difference shown in weak chemotactic chemicals with marginal effect are actually just natural changes in frequency and amplitude, as they do not show statistical significance; the natural variation in oscillation demonstrates the difficulty in reliably measuring the voltage of slime mould. The chemicals with the lowest reliability are very strong repellents, and in some instances, strong attractants, therefore it could be that food substances provide a reliable and repeatable change in frequency while other attractants and repellents impose their chemotaxis by a different method, or are simply lower down the hierarchy of behavioural controls.

In addition to the frequency information provided from the voltage measurement, the amplitude is also recorded, presenting more data than when measured optically. It is reported that the natural oscillating amplitude of a *Physarum Polycephalum* protoplasmic tube is approximately 5 millivolts, while the range for peak to trough amplitude was from 0.25 to 1 millivolts; the most probable explanation for this was that the tubes were residing on top of agar blobs which were placed atop the conductive tape, creating a higher resistance and therefore lower voltage measurement. The amplitude of oscillation may also be a function of the protoplasmic tube length, thickness and state of migration, as briefly mentioned by Kakaiuchi and Ueda [42]. Figure 5 proves that while the frequency is changed, there is a more reliable and statistically significant change in amplitude, however there does not appear to be a trend relating to attractant or repellent strength; it is possible that certain chemicals or chemical types could affect both amplitude and or frequency dependant on their method of action.

While it was possible to measure the absolute changes from before and after frequency, relative changes were calculate, as this would somewhat normalise the data, as each starting frequency varied therefore the magnitude of difference was largely dependant on the frequency before exposure to the chemical; this relative change calculation was also done for the amplitude as the variation of oscillating amplitude varied greatly, presumably due to the migration state and levels of food source at the organism's current position.

At the point of chemical addition there was often a large spike and noisy signal, marked by iii in figure 2, this is due to the disturbance when opening the lid of the Petri-dish and is not a chemical detection signal. The lids were kept on so as to avoid disturbance anomalies in the signal caused by air flow over an open Petri observed by the authors. Adamatzky has recently reported that *Physarum Polycephalum* is touch sensitive [21, 22], and that mechanical stimulation to the surface of the plasmodia results in a large spike in measured electrical potential; it is believed that the spikes which appear just before the chemical is added, is the result of indirect mechanical stimulation transmitted through the Petri-dish lid; the spikes are only acute and do not appear to affect the ensuing oscillation.

The time it takes for the frequency to change is very short; often within one period of oscillation the frequency has changed, and from there onwards the period has changed for the foreseeable future, presumably to change its behaviour towards or away from the chemical. This is very significant, as while it takes an hour to grow a few millimetres, it takes less than 5 minutes for the change in frequency to be instigated; the practical outcome for Physarum computing is significant indeed. It has been shown in this paper that chemotaxis can be measured within 5 minutes rather than having to wait for the organism to grow towards or away from a chemical, a process which can take hours.

The results of the agar attractant experiment show that 2% agar is not a strong attractant, as it only attracted the Physarum in 22% of the time. The 6% agar is more of an attractant than just pure water, due to the frequent migration from a pure water source to a 6% agar gel, supporting the hypothesis that at certain concentrations, agar can be an effective attractant. There appears to be a threshold for attraction of agar between 2% and 6% gel; concentration-based attractant thresholds have been noted in previous literature [3,5]. Transient observations that Physarum grows towards 6% agar have been confirmed with this experiment, supporting the hypothesis that the agar instead provides attraction to Physarum, presumably as agar contains the two polysaccharides, agarose and agaropectin, at least one of which providing a carbohydrate source, with a stronger concentration of agar providing stronger attractant qualities. Testing the electrical response of Physarum to agar showed that 8% agar was as strong an attractant as the strongest VOC attractant, Farnesene; the amplitude was not statistically different.

The variation in frequency and amplitude varies between each chemical tested; this shows that, with a large enough database, the experimental set up as shown here, could be modified slightly and could produce a local non-contact chemical sensor system which would act like a Physarum-based nose. This chemical detector could differentiate between different chemicals or chemical groups without the knowledge of what chemical was present nearby, instead analysing the change in frequency and amplitude when bought into detection range. No attempt to measure the maximum detection range of chemicals was made and it is suggested that different chemicals could have further ranges than others, with concentration dependence also a probable factor. This work demonstrates for the first time, the ability of Physarum to detect individual chemicals over a distance of several centimetres by the relative magnitude and direction of frequency and amplitude change after the chemical is added. One reason for not placing the chemical on an agar gel in this set up is to demonstrate the ability of *P. polycephalum* to detect chemicals in the air; it has been suggested in previous literature that chemotactic chemicals diffuse through the agar medium where *P. polycephalum* can detect them, and while this may be true, it has been proved here that *P. polycephalum* possesses the ability to detect chemicals in the air, as diffusion through the plastic Petri-dish is not possible. Biosensors which act as noses have been developed using in-tact olfactory cells and complex supporting equipment [43], but they are significantly more complicated due to the cell maintenance and hardware requirements.

It is evident that like fungi, Physarum could be used in cell based sensors, as it has the ability to detect a wide range of naturally occurring VOCs as well as simple carbohydrates [2, 3, 4, 5, 9] without the need for genetic modification. Slime mould can live in a variety of conditions so would be suitable for a testing situation where variable conditions were expected. It is known that slime moulds are able to sense oxygen [44], changing their metabolism and life cycle dependant on oxygen, nutrient and light levels, therefore it would be plausible to measure the Physarum cell in response to levels of oxygen in a sample of water, over 5 days at $20^{\circ}C$, and equate this to the BOD sensors, both cell-based and traditional.

Physarum can alternatively be used as a chemical sensor to detect specific chemical or set of chemicals similar to the yeast based biosensors developed by Racek et al [29], detecting glucose, however the range of chemicals that Physarum responds to is significant. It has been demonstrated previously, and indeed in this paper that carbohydrates and VOCs are detected by Physarum. A very interesting outcome of this research is firstly the ability of the cell to detect the chemical without being in direct contact, but also the different responses to chemicals, therefore it would be possible to detect several different chemicals as the organism changes both amplitude and frequency to varying degrees dependant on the chemical. The apparatus costs for this chemical sensor are minimal, the cost of a single Petri-dish, approximately 15 centimetres of conductive tape and a high resolution voltage data logger; the cost of the latter equipment could be replaced with dedicated electronics circuit to significantly reduce the cost. The amplitude and frequency change was calculated manually in this instance, however could be replaced by simple software to interpret and analyse the voltage output. The range of chemicals that Physarum could detect is far beyond any that have been tested thus far by everyone that has done so, meaning that the potential for chemical sensors utilising Physarum as the cellular contributor is vast. If Physarum can be genetically modified, it too could form the base of in vitro testing, replacing bacteria with a more suitable and similar eukaryotic cell, indeed it may be not need genetic modification in some instances. The known Physarum attractant, glucose, present in blood at varying levels, could be developed into an assay to determine blood glucose in blood of diabetic patients. It is evident from this work that *P. polycephalum* could be a suitable alternative to both bacterial and fungal biosensors; a comparative list, table 2, is drawn below of advantages and disadvantages for all three types of biosensor, showing the practical limitations such as shelf life, fragility and variety of substances which can be detected by the same sensor, as well as cost and genetic modification; epithelial cell based biosensors have a longevity of up to 37 days [45], an improvement on the bacterial cells, but with Physarum, with refreshing conditions is it possible to grow Physarum for many months or dry it out for years with reanimation at a later date. Table 2 shows that Physarum is certainly comparable with bacterial and fungal biosensors, and this paper proves the concept of a Physarum biosensor; while most bacterial and fungal sensors can detect single or on occasion a small number of chemicals [Baronian 04, 46], Physarum can detect and differentiate 8 chemicals and in the case of Agar, also detect concentrations.

Table 2. Comparison of Bacteria, Fungus and Physarum Bio-sensors.

|  | **Bacterial** | **Fungal** | **Physarum** |
|---|---|---|---|
| **Shelf-Life** | 2 weeks [46] | 1 year [25] | Several years |
| **Operational life** | 5 years [47] | 2 months [48] | Unknown, presumed several months |
| **Method of growth substrates** | Immobilised, Suspension. [41] | Immobilised, Membrane or Agar entrapment, Suspension. [25] | Glass, Metals, Plastics, Agar. [19] |
| **Number of chemicals detected** | Large [49] | Large [25] | Unknown but large |
| **Signal detection methods** | Optical, Amperometric, Voltametric. [41, 50] | Amperometric, pH, $CO_2$, Optical, Growth rate. [25] | Voltage, Current, Growth pattern/Optical. [19] |
| **Genetic modification** | Well established | Established | Transient only [28]. |

The current iterations of Physarum computers use the growth and migration of the plasmodia as the output [13, 31, 51], but it is now evident that the electrical output could be used as a much quicker and simpler method of calculation, offering a significant advancement in the field. For the application of Physarum computers, investigating figure 3 shows that for a single stimulus, the combination of amplitude and frequency can be used to differentiate chemicals from one another.

While no concurrent optical and electronic potential measurements of the oscillations in *P. polycephalum* were made due to the difficulty of simultaneously measuring and observing several experiments and the significant manual visual processing required, it is now safe to suggest that the oscillations recorded optically are very closely related if not identical to those recorded electrically. The demonstration that, in most circumstances, known attractants increased the oscillation frequency while repellents decreased it, as shown in both previous literature as measured optically, and in this paper measured electrically, are in agreement, it is a very strong case for suggesting that membrane movement is at least directly correlated to if not governed by the electrical potential of the membrane. The movement of ions through voltage gated ion channels has long been known as the source of muscle movement, and with the theory that *P. polycephalum* membrane movements are controlled with actin-myosin type contractions [1,4, 52], which are themselves controlled by voltage gated ion channels, lends stronger evidence to the correlation between optically and electrically measured oscillation.

Other chemicals may trigger a non-specific receptor for attractants or repellents or simply disrupt the plasmodium membrane leading to decreased localised oscillation resulting in migration away from the chemical, as other sections of the membrane on the opposite side of the plasmodium continues oscillation hence migration and or growth. It has been suggested that some VOCs are known to bind to certain receptors such as Limonene which binds to Adenosine A(2A) receptors and increase cAMP and calcium concentration [9], may explain why certain chemicals have chemotactic properties. Without knowing all the receptors on *P. polycephalum's* membrane, it is not totally clear how the chemicals induce their chemotactic response; further work on the membrane receptors and chemicals will shed light on this area. The behaviour of symbiotic or even parasitic-like defence from other plants may appear as intelligence, however it is more likely the evolution of a certain receptor which is sensitive to a chemical which deters predators has ensured survival of the organism, so the single celled *P. polycephalum* has no decision making skills or brain to speak of, instead is it governed by a set of behavioural qualities such as finding a food source and moving towards favourable conditions, which make it ideal for organism or bio-inspired computing.

## 5. Conclusion

It has been demonstrated in this work that the frequency change for attractants and repellents is the same as in previous literature, being that chemicals which impose positive chemotaxis in *P. polycephalum* increase the frequency of oscillations along a protoplasmic tube while negative chemotactic chemicals reduce the frequency of oscillations when measured electrically. Chemicals that appear neutral to the organism are still detected, manifested by a change in amplitude of the oscillation. *P. polycephalum* can identify individual chemicals by the relative amplitude and frequency change after exposure; the detection of chemicals can occur at 4 centimetres, without diffusion through a growth medium such as agar. It is possible that this set up could be employed as a chemical sensor, allowing the contactless detection of VOCs as well as potentially other chemicals.

Agar has been shown to have attractant or positive chemotactic properties with respect to the slime mould, with increasing concentrations exaggerating the effect; future investigations with *P. polycephalum* attractants and repellents should note that concentrations above 2% non-nutrient agar still show mild positive chemotaxis which may interfere with results.

This paper has documented work which enhances the understanding of the protoplasmic streaming and chemotaxis, for the first time electronically measuring the frequency and amplitude change of chemotaxis. The chemotactic effect occurs rapidly and can be measured using this technique, which has the potential to enhance Physarum computing from visually observed growth which may take days, to electronically measured response occurring within a matter of minutes. The ability of *P. polycephalum* to detect chemicals and also differentiate between them has the potential to be used as a chemical sensor which can detect a variety of different chemicals. In addition to having developed the proof of concept Physarum chemical sensor, it is believed that this knowledge will allow the development of the next generation of efficient Physarum computers.